\documentclass[twocolumn,iop, aps,superscriptaddress,showpacs]{revtex4-1}

\usepackage{mathptmx}
\usepackage{subfigure}
\usepackage{dcolumn}
\usepackage{amsmath,amssymb}
\usepackage{bm}
\usepackage{color}
\usepackage{latexsym}
\usepackage{epstopdf}
\usepackage{color}
\usepackage[english]{babel}
\usepackage{latexsym}

\usepackage{psfrag,graphicx} 
\usepackage{epsf} 
\usepackage{subfigure} 
\usepackage{amsmath} 
\usepackage{amssymb} 
\usepackage{amsfonts}
\usepackage{bm}
\usepackage{natbib}
\usepackage{epstopdf}
\usepackage{appendix}

\definecolor{mygrey}{gray}{0.35}
\definecolor{myblue}{rgb}{0.2,0.2,0.8}
\definecolor{myzard}{cmyk}{0,0,0.05,0}
\definecolor{mywhite}{rgb}{1,1,1}
\definecolor{myred}{rgb}{1,0.,0.3}

\usepackage[colorlinks=true,citecolor=myblue,linkcolor=myred]{hyperref}

\def\be{\begin{equation}}
\def\ee{\end{equation}}
\def\ba{\begin{align}}
\def\enda{\end{align}}
\def\bi{\begin{itemize}}
\def\ei{\end{itemize}}

 \def\ee{\mathord{\rm e}}

 \def\ee{\mathord{\rm e}}

\renewcommand{\ee}{{\rm e}}

\def\beq{\begin{equation}}
\def\beq{\begin{equation}}
\def\eeq{\end{equation}}

\begin{document}

\title[Short Title]{Multi-Qubit Gate with Trapped Ions for Microwave and Laser-Based Implementation}

\author{I. Cohen}
\affiliation{Racah Institute of Physics, The Hebrew University of Jerusalem, Jerusalem 91904, Givat Ram, Israel,}

\author{S. Weidt}
\affiliation{Department of Physics and Astronomy, University of Sussex, Brighton BN1 9QH, United Kingdom}

\author{W. K. Hensinger}
\affiliation{Department of Physics and Astronomy, University of Sussex, Brighton BN1 9QH, United Kingdom}

\author{A. Retzker}
\affiliation{Racah Institute of Physics, The Hebrew University of Jerusalem, Jerusalem 91904, Givat Ram, Israel,}


\begin{abstract}
{A proposal for a phase gate and a M{\o}lmer-S{\o}rensen (MS) gate in the dressed state basis is presented. In order to perform the multi-qubit interaction, a strong magnetic field gradient is required to couple the phonon-bus to the qubit states. The gate is performed using resonant microwave driving fields together with either a radio-frequency (RF) driving field, or additional detuned microwave driving fields. The gate is robust to ambient magnetic field fluctuations due to an applied resonant microwave driving field. Furthermore, the gate is robust to fluctuations in the microwave Rabi frequency and is decoupled from phonon dephasing due to a resonant RF or a detuned microwave driving field. This makes this new gate an attractive candidate for the implementation of high-fidelity microwave based multi-qubit gates. The proposal can also be realized in laser-based set-ups.
}
\end{abstract}
\maketitle

High-fidelity quantum gates are a crucial element in the growing field of quantum information processing (QIP) \cite{Quantum Information}. Many theoretical proposals for quantum entangling gates have been considered for trapped ions \cite{gate proposal 1,gate proposal SM1,gate proposal SM2, gate proposal GEOMETRIC,gate proposal MONROE1,gate proposal MONROE2,gate proposal ROOS,gate proposal MICROWAVE,gate proposal BERMUDEZ1,gate proposal BERMUDEZ2}, one of the most promising candidates for QIP  \cite{QIP candidate 1,QIP candidate 2,QIP candidate 3}. These proposals have triggered impressive experimental realizations  \cite{gate realization DIDI,CZ_gate,Monroe1995,gate realization BERMUDEZ,Wilson,gate realization MICROWAVE,gate realization WUNDERLICH1,Ozeri_gate,blatt_gate,monroe_gate,lucas}. Although these experiments within the laser based designs have achieved high fidelities \cite{lucas,gate realization BERMUDEZ,Ozeri_gate,blatt_gate,monroe_gate}, the achieved fidelities within the microwave based designs have been limited. 


Considerable theoretical efforts have been made to counter the fidelity-damaging effects. 
Techniques, such as dynamical decoupling using echo-pulse sequences \cite{DD1,DD2}, and its continuous version using a continuously applied driving field \cite{DD3,DD4,DD5,DD6,Wang1,Nati}, have been proposed and realized experimentally \cite{gate realization WUNDERLICH1,WINNI,John_dec1,John_dec2}. Recently, a combination of the continuous techniques with gate operators has been proposed \cite{gate proposal BERMUDEZ1,gate proposal BERMUDEZ2} and realized \cite{gate realization BERMUDEZ,Wilson} for the laser-induced implementations. 

In this manuscript, we introduce a scheme for a geometric gate $\sigma_z\otimes\sigma_z$, and a M{\o}lmer-S{\o}rensen (MS) gate $\sigma_x\otimes\sigma_x$ in the dressed state basis for microwave-based implementations. It can also be implemented in laser-based experimental set-ups. Our scheme combines the gate operator and continuous dynamical decoupling which results in the gate being decoupled from the main fidelity damaging noise sources such as ambient magnetic field and Rabi frequency fluctuations. 

In what follows, we derive the gate Hamiltonian for a microwave-based design, while highlighting the dynamical decoupling processes which protect from the main noise sources. We then go on showing explicitly how our scheme could also be implemented using laser rather than microwave radiation.

\section{method}

In the microwave-based implementation the required two spin states defined as $\left\lbrace  \left\vert 1 \right\rangle, \left\vert 0 \right\rangle   \right\rbrace$ are separated by an energy splitting in the microwave regime \cite{gate realization MICROWAVE,gate realization WUNDERLICH1,WINNI}. The spin-spin interaction, which constructs multi-qubit gates, is induced via the exchange of virtual phonons between the ions, in the weak coupling regime \cite{gate proposal SM1}, or real phonons in the strong coupling regime \cite{gate proposal SM2,gate proposal ROOS}. Therefore, a coupling between the spin and the interaction-mediating phonon is needed. Due to the negligibly small Lamb-Dicke parameter associated with microwave radiation, inducing a sufficiently strong spin-phonon coupling requires the use of a large magnetic field gradient.

There are two basic approaches to obtain a magnetic field gradient. One approach makes use of an oscillating magnetic field gradient induced by the microwave driving field in the near field regime, where the spin-phonon coupling is already linked to the driving fields which perform the gate \cite{gate proposal MICROWAVE,gate realization MICROWAVE}. This enables to utilize qubits that can have first order magnetic field insensitive transitions \cite{lucas}; namely, the qubit states should be magnetic field sensitive, yet the energy difference between the two states is insensitive to the first order in the magnetic field. Therefore, these qubits are decoupled from the ambient magnetic noise, and thus can be considered as clock qubits.

The other approach uses a static magnetic field gradient \cite{MAGIC1,MAGIC2}. Here, the qubits must have first order field sensitive transitions, making them extremely sensitive to the ambient magnetic noise, which imposes the use of dynamical decoupling techniques.  Furthermore, in this approach the spin-phonon coupling is not linked to driving field transitions that eventually perform the gate. In order to analyze this setting, one can apply a polaron-like transformation \cite{Wolff}. In that case, there always remain undesired transitions, which originate directly from the static magnetic field gradient \cite{cohen,Gatis}. Not accounting for these transitions may damage the gate fidelity, and adiabatically eliminating them imposes restrictions on experimental parameters, such as an upper bound to the effective Lamb-Dicke parameter, which can significantly increase the gate duration. Instead of using the dressing fields as the gate performing term, in this proposal, the gate is generated directly by the static magnetic field gradient.

To realize a gate in the strong coupling regime, it is necessary to have sideband addressing of the resolvable secular frequencies, namely to couple to only one common mode of motion of secular frequency $\nu$. This is done by applying a microwave driving field on-resonance with the bare qubit $\left\lbrace  \left\vert 1 \right\rangle, \left\vert 0 \right\rangle   \right\rbrace$, whose Rabi frequency is nearly equal to the selected common mode of motion, $\Omega=\nu -\epsilon$, and far detuned from all the other modes. For a two-qubit gate, an increase of the gate speed could be obtained by coupling to two modes, with opposite detunings. This is done by setting the microwave Rabi frequency exactly between the two modes; however, for the simplicity of the derivation, we only consider coupling to one mode. 


\section{derivation of the Hamiltonian}
\begin{figure}
   \centering
  \includegraphics[width=0.5\textwidth,natwidth=1280,natheight=720]{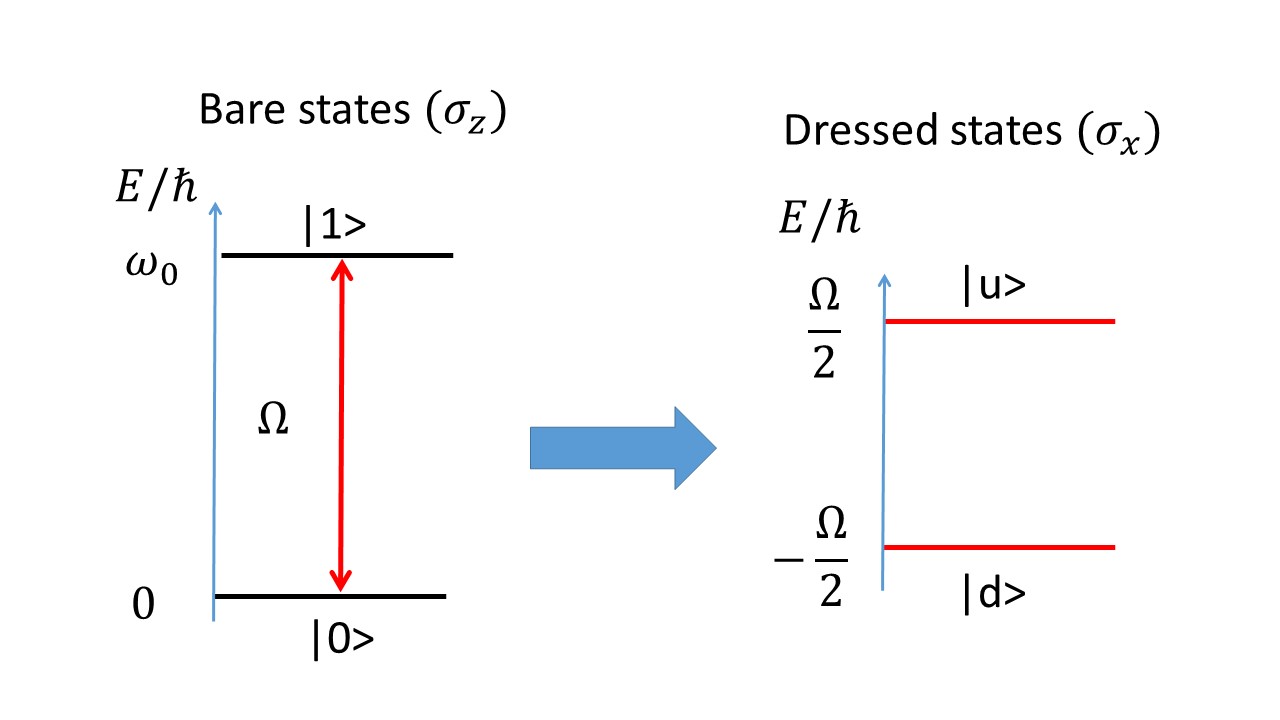}
  \caption{{\bf Moving from the bare states to the dressed states.} In the microwave-based designs, due to the magnetic gradient, each ion feels different magnetic field, and therefore, it has a different bare state energy splitting $\sum_j\frac{\omega_0^j}{2}\sigma_z^j$. By applying microwave driving fields on resonance with the bare state energy splitting of the different ions, we move to the dressed state basis, in a perpendicular direction $\frac{\Omega}{2}\sum_j\sigma_x^j$. The energy gap in the perpendicular direction protects against the magnetic noise.
  \\
  In the laser-based designs, if the qubit transition frequency is in the optical regime, instead of the resonant microwave driving fields, we may apply a single resonant laser transition, in order to move to the dressed states.}
    \label{bare}
\end{figure}

 Due to the constant magnetic gradient and the  difference in the  magnetic moments of the spin states $\left\vert 0 \right\rangle$ and $\left\vert 1 \right\rangle$, a state-dependent force is obtained.
 In addition, each ion feels different magnetic field, and as consequence, different Zeeman splitting. Therefore, the Hamiltonian can be written as
\begin{equation}
\begin{split}
H=\nu b^\dagger b   +\sum_j\left(\frac{\omega_0^j}{2}\sigma^j_z +\frac{\nu\eta^j}{2}( b^\dagger +b) \sigma^j_z   +\Omega\sigma^j_x \cos \omega_0^j t\right),
\end{split}
\label{start}
\end{equation}
which includes the selected vibrational mode with secular frequency $\nu$, the bare state energy structure of the $j^{th}$ ion, the magnetic field gradient, and resonant microwave driving fields applied to the $j^{th}$ ion respectively (fig. \ref{bare}). The Pauli spin matrices $\sigma_z = \left\vert 1 \right\rangle \left\langle 1 \right\vert - \left\vert 0 \right\rangle \left\langle 0 \right\vert$, $\sigma_x = \left\vert 1 \right\rangle \left\langle 0 \right\vert + \left\vert 0 \right\rangle \left\langle 1 \right\vert$, $b^\dagger$ and $b$ are the phonon creation and annihilation operators, respectively, and $\eta^j=g\mu_B \partial_z B Z^j/\sqrt{2m\nu^3}$ is the effective Lamb-Dicke parameter, with $Z^j$ being the selected normal mode coefficient. For simplicity we assume that $Z^j$ is the same for all the ions, namely we are coupled to the center of mass mode. 


Large magnetic gradient enables single addressing \cite{large gradient}, where a microwave driving field that is on resonance with one ion transition, is far detuned from the other ion transitions. The neglected off-resonant driving fields contribute an A.C Stark shift which results in an undesirable phase shift, and will be discussed in the next section.

According to the microwave driving field's initial phase, the dressed states are determined. For simplicity, in the interaction picture we consider a vanishing initial phase, such that the dressed state basis is the eigenstates of $\sigma_x$, namely $\left\vert u \right\rangle = \left( \left\vert 1 \right\rangle+\left\vert 0 \right\rangle   \right)/\sqrt{2} $, $\left\vert d \right\rangle = \left( \left\vert 1 \right\rangle   -   \left\vert 0 \right\rangle   \right)/\sqrt{2} $ (\ref{bare}). 

Moving to the interaction picture with respect to the microwave energy splitting, $\sum_j(\omega_0^j/2)\sigma^j_z$, followed by transforming to the dressed state basis, the operators are transformed as follows:
$\sigma_x \rightarrow S_z$,$\sigma_y \rightarrow S_y$, and $\sigma_z  \rightarrow -S_x$, by where $S_\alpha$ are the Pauli matrices in the $\alpha$ direction, in the dressed state basis. Thus, the Hamiltonian of eq.\ref{start} is transformed and can be written as
\begin{equation}
H_I=\nu b^\dagger b + \sum_j\left(-\frac{\nu\eta}{2}( b^\dagger +b) S^j_x   +\frac{\Omega}{2}S^j_z\right),
\label{start2}
\end{equation}
after making the rotating wave approximation (RWA), when assuming $\Omega\ll 4\omega_0$.  The neglected fast rotating terms of the microwvae driving fields contribute an A.C Stark shift which results in an undesirable phase shift, and will be discussed in the next section.

\begin{figure}
   \centering
  \includegraphics[width=0.5\textwidth,natwidth=1280,natheight=720]{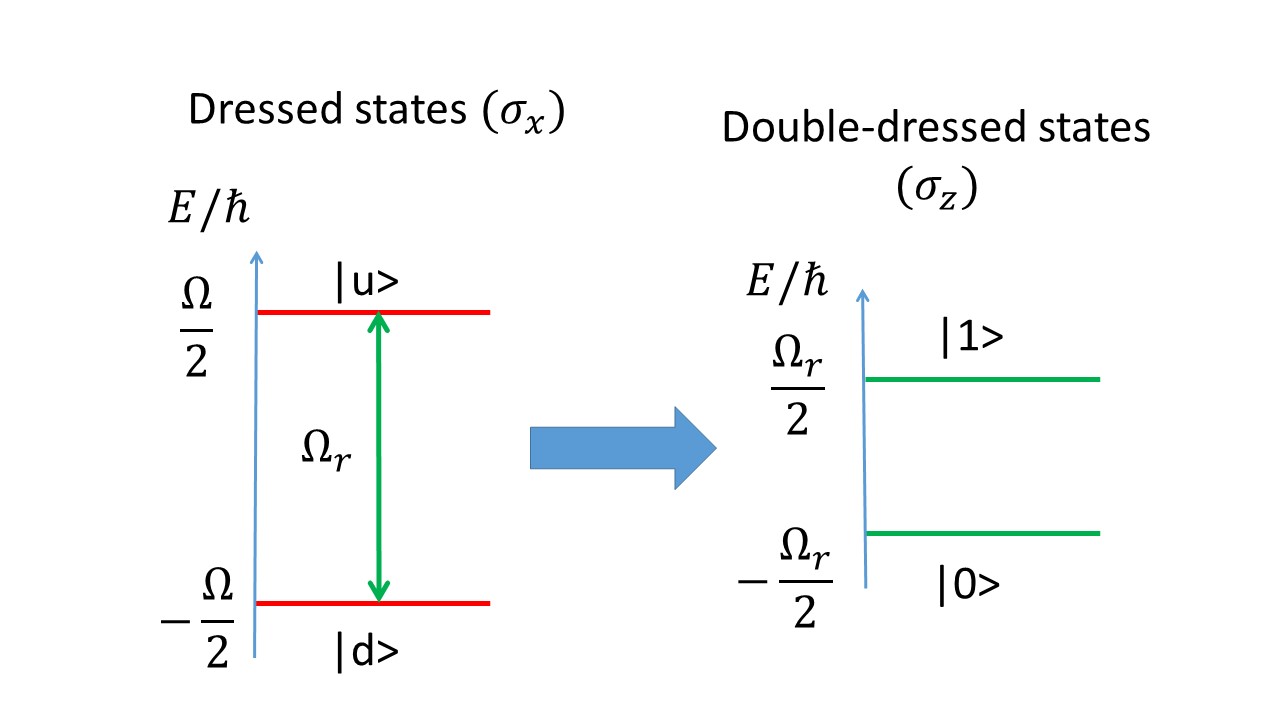}
  \caption{{\bf Moving from the dressed states to the double dressed states using RF field.} By applying a single RF driving field on resonance with the dressed state energy splitting $\frac{\Omega}{2}\sum_j\sigma_x^j$, we move to the double-dressed state basis, in a perpendicular direction $\frac{\Omega_r}{2}\sum_j\sigma_z^j$. The energy gap in the perpendicular direction protects against the noise that originates from the Rabi frequency fluctuations $\Omega$.}
    \label{dressed RF}
\end{figure}

\begin{figure}
   \centering
  \includegraphics[width=0.5\textwidth,natwidth=1280,natheight=720]{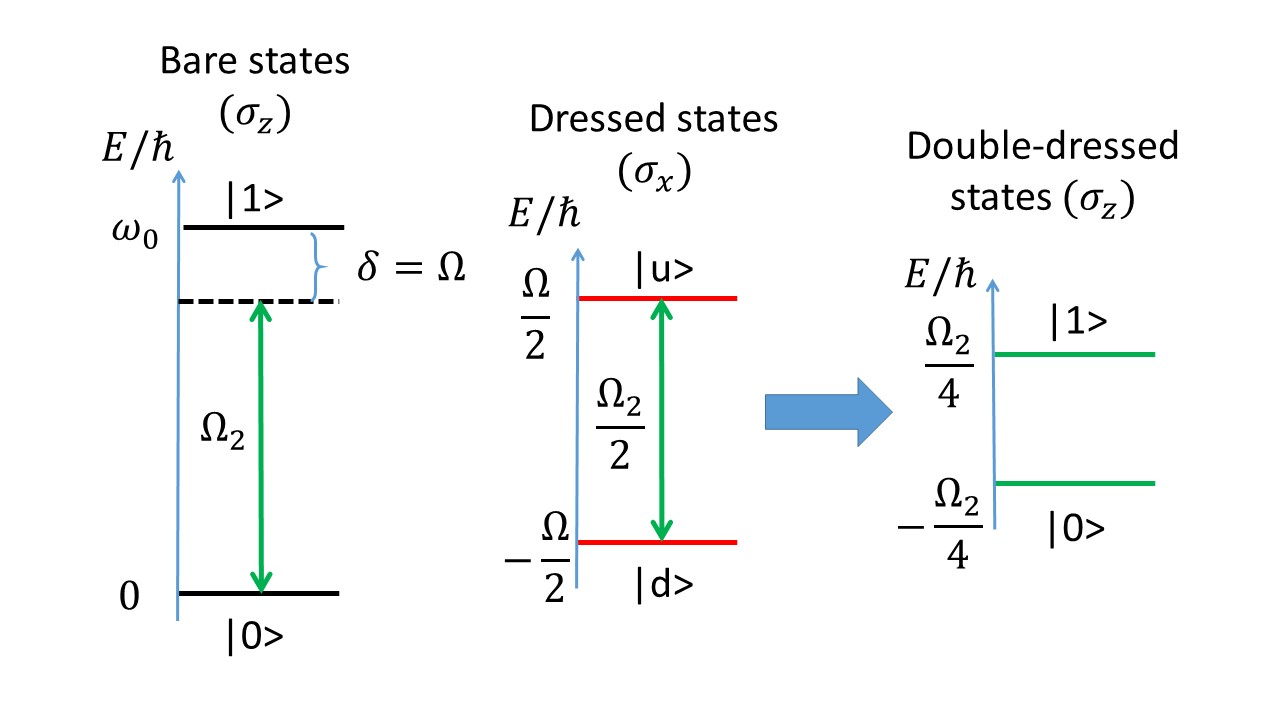}
  \caption{{\bf Moving from the dressed states to the double dressed states using detuned transition in the bare state basis.} In the microwave-based designs, due to the magnetic gradient, each ion feels different magnetic field, and therefore, it has a different bare state energy splitting $\frac{\omega_0^j}{2}\sum_j\sigma_z^j$. Applying another microwave driving fields, which are detuned from the bare state energy splitting by exactly the Rabi frequency of the resonant microwave driving field, $\delta=\Omega$, and phase-locked with it, is effectively equivalent to driving resonant transition with the dressed state energy structure. This takes us to the double-dressed state basis, in a perpendicular direction $\frac{\Omega_2}{4}\sum_j\sigma_z^j$. The energy gap in the perpendicular direction protects against the noise that originates from the Rabi frequency fluctuations $\Omega$.
A single detuned microwave driving field can also be used for the laser-based designs, if the qubit transition frequency is in the microwave regime. If the qubit transition frequency is in the optical regime, instead of the detuned microwave driving field, we may apply a single laser transition, which is detuned from the bare state energy splitting by exactly the Rabi frequency of the resonant laser transition, $\delta=\Omega$. In that way the system moves to the double-dressed states which are decoupled from the Rabi frequency fluctuations of the resonant laser transition.   
  }
    \label{dressed M}
\end{figure}

In the interaction picture with respect to the dressed state energy splitting, $(\Omega/2)\sum_j S^j_z$, the red sideband transition, which originates directly from the magnetic field gradient, can be written as
\begin{equation}
H_{red}=-\frac{\eta\nu}{2}\sum_j\left(  b^\dagger e^{i\nu t} +h.c   \right)  \left(   S^j_+ e^{i\Omega t} +h.c \right).
\label{red}
\end{equation}
In the second-order perturbation approach the red sideband transitions result in a flip-flop Hamiltonian $\sum_{i,j} S^j_+ S^i_- + h.c $ with an undesired AC Stark shift term, which is coupled to the number of phonons $\sum_jS_z^j b^\dagger b$, behaving as a phonon dephasing source. Instead of adiabatically eliminating these transitions, we use them as our gate-performing terms, while suppressing the dephasing term.
Transforming eq. \ref{red} to the desired gate transitions can be achieved in two ways: (a) by applying a single RF driving field (fig. \ref{dressed RF}), which is on-resonance with the dressed state energy splitting determined by the microwave driving field, and polarized in the axial direction, which is determined by the external magnetic field.(b) By applying another microwave driving fields (fig. \ref{dressed M}), which are detuned from the bare state energy splitting by exactly the Rabi frequency of the resonant microwave driving field, $\delta=\Omega$, and phase-locked with it. These additional driving fields can be represented as
\begin{equation}
\begin{split}
(a) H_{rf}=\Omega_r\sum_j\sigma^j_z\cos \Omega t\\
(b) H_{\mu 2}=\Omega_{2}\sum_j\sigma^j_x \cos (\omega^j_0 -\Omega) t
\end{split}
\label{radio1}
\end{equation}
Following the same stages of derivation up to this point, namely, moving to the interaction picture with respect to the qubit splitting, followed by moving to the interaction picture according to the dressed state energy structure, the additional terms (eq.\ref{radio1}a,b) can be written as
\begin{equation}
\begin{split}
(a) H_{rf}'= -\frac{\Omega_r}{2}\sum_j S^j_x \\ 
(b) H_{\mu 2}'=-\frac{\Omega_{2}}{4}\sum_j S^j_x,
\end{split}
\label{radio2}
\end{equation}
after making the RWA, where we assume that $\Omega_{r},\Omega_{2} \ll 4\Omega$. The neglected fast rotating terms contribute an A.C Stark shift which results in an undesirable phase shift. Using detuned microwave fields instead of a single RF field, contribute another A.C Stark shift, due to the neglected off-resonant contribution. These undesired A.C Stark shifts will be discussed in the next section.





Transforming to the double-dressed state basis, and moving to the interaction picture with respect to the time-independent term of the RF driving field or the detuned microwave driving field (eq.\ref{radio2}), the red sideband transitions (eq.\ref{red}) become
\begin{equation}
H_{I}'=\frac{\eta\nu}{4}\sum_j \left(  b^\dagger e^{i\nu t} +h.c   \right)  \left(   \left( \sigma^j_z - \sigma^j_+ e^{i\Omega_r t}  +  \sigma^j_- e^{-i \Omega_r t}  \right) e^{i\Omega t} +h.c \right),
\label{Lamb-Dicke regime}
\end{equation}
where for simplicity we have merged the notations: $\Omega_r=\Omega_{ 2}/2$. 

By setting $\epsilon:=\nu-\Omega \ll \Omega_r$, the only transition which is not suppressed, and thus survives, is the desired phase gate transition in the double-dressed state basis, or the MS gate transition in the dressed state basis
\begin{equation}
H_{gate}=\frac{\eta\nu}{4}\sum_j\sigma^j_z\left(b^\dagger e^{i\epsilon t} +b e^{-i\epsilon t} \right)=-\frac{\eta\nu}{4}\sum_jS^j_x\left(b^\dagger e^{i\epsilon t} +b e^{-i\epsilon t} \right).
\label{red2}
\end{equation}

In order to obtain the effective Hamiltonian of the gate, we can use the Magnus expansion \cite{MAGNUS1} of the time propagator corresponding to the time-dependent Hamiltonian shown in eq.\ref{red2}. The Magnus expansion can then be written as
\begin{equation}
\begin{split}
U(t)=\exp( - i\int_0 ^t dt' H_{gate}(t')\\
 - \frac{1}{2} \int_0 ^t  dt' \int_0^{t'}  dt''  \left[ H_{gate}(t'),H_{gate}(t'')  \right]  +... ).
\end{split}
\label{magnus}
\end{equation}
For this specific Hamiltonian, only the first two orders of the Magnus expansion contribute, whereas higher orders vanish  \cite{MAGNUS2}. An alternative approach to obtain the same result is to follow the MS gate notation in Ref. \cite{gate proposal SM2}. According to it, the gate Hamiltonian can be represented by 
\begin{equation}
H_{gate}=\frac{\eta\nu}{\sqrt{8}} \sum_j\sigma^j_z\left( x\cos\epsilon t+p\sin\epsilon t\right)=:f(t)x+g(t)p.
\label{SM}
\end{equation}
where $x = (b^\dagger +b)/\sqrt{2}$ and $p = -i(b^\dagger -b)/\sqrt{2}$ are the dimensionless operators of the position and momentum respectively.
Since the first two terms of eq.\ref{SM} commute with their mutual commutator, we obtain the gate's exact time propagator given by
\begin{equation}
U(t) = e^{-iA(t)}e^{-iF(t)x}e^{-iG(t)p},
\end{equation}
where 
\begin{equation}
F(t)=\int_0^t f(t')dt'=\frac{1}{\sqrt{8}}\frac{\eta\nu}{\epsilon}\sum_j\sigma^j_z\sin\epsilon t 
\label{motion coupling1}
\end{equation}
\begin{equation}
G(t)=\int_0^t g(t')dt'=\frac{1}{\sqrt{8}}\frac{\eta\nu}{\epsilon}\sum_j\sigma^j_z\left( 1- \cos\epsilon t\right)
\label{motion coupling2}
\end{equation}
\begin{equation}
A(t)=-\int_0^t F(t')g(t')dt'=-\frac{1}{16}\frac{\eta^2\nu^2}{\epsilon}\sum_{i,j}\sigma_z^i\sigma_z^j\left( t-\frac{1}{2\epsilon} \sin 2\epsilon t\right).
\label{Agate}
\end{equation}
In order to obtain the required behavior of the gate, we have two restrictions: First, we should suppress the entanglement of the spin to the motion, that is to say, we set $F(\tau)=G(\tau)=0$, by choosing $\tau=2\pi K / \epsilon$, with an integer $K$, representing the integer number of circles in phase space. Second, we would like to have a maximally entangled state when starting with a separable state, so we require $A(\tau)=-(\pi/8)\sigma_z^i\sigma_z^j$. These constraints determine the connection between the detuning and the Rabi frequency of the sideband transitions, and the gate time, $\epsilon=\eta\nu\sqrt{K}$ and  $\tau=2\pi\sqrt{K}/\eta\nu$, respectively. 

All of the assumptions made during the above derivation can be summarized as
\begin{equation}
\eta\nu/4\sqrt{K} = \epsilon/4 \ll \Omega_r/4 \ll \nu \thicksim \Omega \ll 4\omega_0.
\label{assumption}
\end{equation}

\section{Calculating the neglected terms}

As was mentioned above, the neglected off-resonant microwave driving fields in eq. \ref{start}, 
and the fast rotating terms of the second driving field (the RF field or the additional detuned microwave driving fields) in eq. \ref{radio2} contribute an A.C Stark shift which results in an undesirable phase shift. In addition, except from the gate operator eq. \ref{red2}, we have also neglected other terms from eq. \ref{Lamb-Dicke regime}. In this section we will estimate the leading  contribution of these fidelity damaging terms, and suggest a way to counter them using spin-echo. 


large magnetic gradient enables single addressing \cite{large gradient}, since each ion feels different magnetic field, and as consequence, a different Zeeman splitting. 
Suppose we have two ions in our trap, aligned in the $z$ direction. Thus, the energy difference between the two ions is
\begin{equation}
\Delta\omega_0=\omega_0^{1}-\omega_0^{2}=g\mu_b\partial_z B \Delta Z,
\end{equation}
where 
\begin{equation}
\Delta Z=\left(\frac{2e^2}{4\pi\epsilon_0 M \nu^2}\right)^{1/3}
\end{equation}
is the distance between the ions in the axial direction, $\epsilon_0$ is the electric permittivity, $M$ is the ion mass, and $e$ is the elementary charge. Taking the off-resonant microwave driving fields into account, another term appears in the rotating frame of the bare energy structure
\begin{equation}
\frac{\Omega}{2}\left( \sigma_+^1 e^{i\Delta\omega_0 t}+\sigma_+^2 e^{-i\Delta\omega_0 t}\right) +h.c.
\label{delta}
\end{equation}

We have dropped the fast rotating terms of the second driving field (the RF field or the additional detuned microwave driving fields). Therefore, in eq. \ref{radio2}, for the RF driving case, we should have additional rotating terms: 
\begin{equation}
-\frac{\Omega_r}{2}\left(S_+^1+S_+^2\right) e^{2i\Omega t} +h.c.
\label{fast RF}
\end{equation}


In addition, in the previous section we have also neglected the following terms from eq. \ref{Lamb-Dicke regime}:
\begin{equation}
\frac{\eta\nu}{4}\left(  b^\dagger e^{i\nu t} +h.c   \right)  \left( \sum_j  \left( - \sigma^j_+ e^{i\Omega_r t}  +  \sigma^j_- e^{-i \Omega_r t}  \right) e^{i\Omega t} +h.c \right),
\label{XY}
\end{equation}
and 
\begin{equation}
\frac{\eta\nu}{4}\sum_j\sigma^j_z\left(  b^\dagger e^{i(\nu+\Omega) t} +h.c   \right). 
\label{ZZ} 
\end{equation}

We can evaluate the leading contributions of these terms by taking into account the time-independent terms of Magnus expansion. For eq. \ref{delta},\ref{fast RF}, the leading contribution comes from the second order of Magnus expansion, and gives the following A.C Stark shift:
\begin{equation}
-\Omega_r\left[3\left(\frac{\Omega}{4\Delta}\right)^2+\frac{1}{2}\left(\frac{\Omega_r}{4\Omega}\right)^2\right]\sum_j\sigma_z^j.
\label{AC1}
\end{equation} 
The leading contributions of eq. \ref{XY} are:
\begin{equation}
\frac{2}{\Omega_r}\left(\frac{\nu\eta}{4}\right)^2 \sum_j\sigma_z^j \left(b^\dagger b +\frac{1}{2}\right),
\label{AC2}
\end{equation}
which is an A.C Stark shift also coupled to phonons, and
\begin{equation}
2\left(\frac{\nu\eta}{4}\right)^2\left( \frac{\epsilon}{\Omega_r^2} -\frac{1}{\nu+\Omega} \right)\left( \sigma_+^1\sigma_-^2+h.c  \right),
\label{XY1}
\end{equation}
which couples between the two ions. Another coupling term comes from 
the leading contribution of eq. \ref{ZZ}:
\begin{equation}
-2\left(\frac{\nu\eta}{4}\right)^2 \frac{1}{\nu+\Omega}\sigma_z^1\sigma_z^2.
\label{ZZ1}
\end{equation}

The A.C Stark shifts of eq. \ref{AC1} and \ref{AC2} cause dephasing, and damage the gate fidelity. 
In order to compensate for the dephasing A.C Stark shifts, we can move to the interaction picture with respect to these terms, by applying a global rotation about the $z$ axis. Alternatively, this undesired contribution can be compensated for by 
applying a single $\pi$-pulse after half of the gate time. Note that only when an integer number of cycles in phase space have been completed, should a $\pi$-pulse be taken; therefore, a $\pi$-pulse in the middle of the gate determines the total number of cycles, $K$, to be even.

Instead of applying a $\pi$-pulse, we can use an alternative procedure, by applying a $\pi$-phase flip of the RF driving field, in the middle of the gate \cite{gate proposal ROOS, pi}. This is due to the following three reasons: 1.) Applying a $\pi$-phase flip of the RF driving field, does not change the gate Hamiltonian (eq. \ref{red2}). 2.) It turns out that all the leading contributions of the A.C Stark shifts (eq. \ref{AC1},\ref{AC2}) are an odd function of $\Omega_r$, therefore, changing the sign of $\Omega_r$, would change the sign of the undesired terms. 3.) These A.C Stark shifts also commute with the gate Hamiltonian (eq. \ref{red2}), and therefore, one dynamical-decoupling-like $\pi$-phase flip, applied in the middle of the gate, is sufficient to counter the undesired A.C Stark shifts. Another advantage of applying a $\pi$-phase flip of the RF driving field is that it can work even for $K=1$, as opposed to the $\pi$-pulse technique.

Since eq. \ref{AC2} is also coupled to the phononic states $b^\dagger b$, which are populated during the gate duration, (especially due to the electric field effect, as will be discussed below),  applying a single $\pi-$phase flip may not be sufficient to obtain high fidelity. For that reason, we can apply a sequence of $\pi-$phase flips.  Note that this sequence can be performed even within a single cycle in phase plane $(K=1)$. 

\begin{figure}
   \centering
  \includegraphics[width=0.5\textwidth,natwidth=1921,natheight=964]{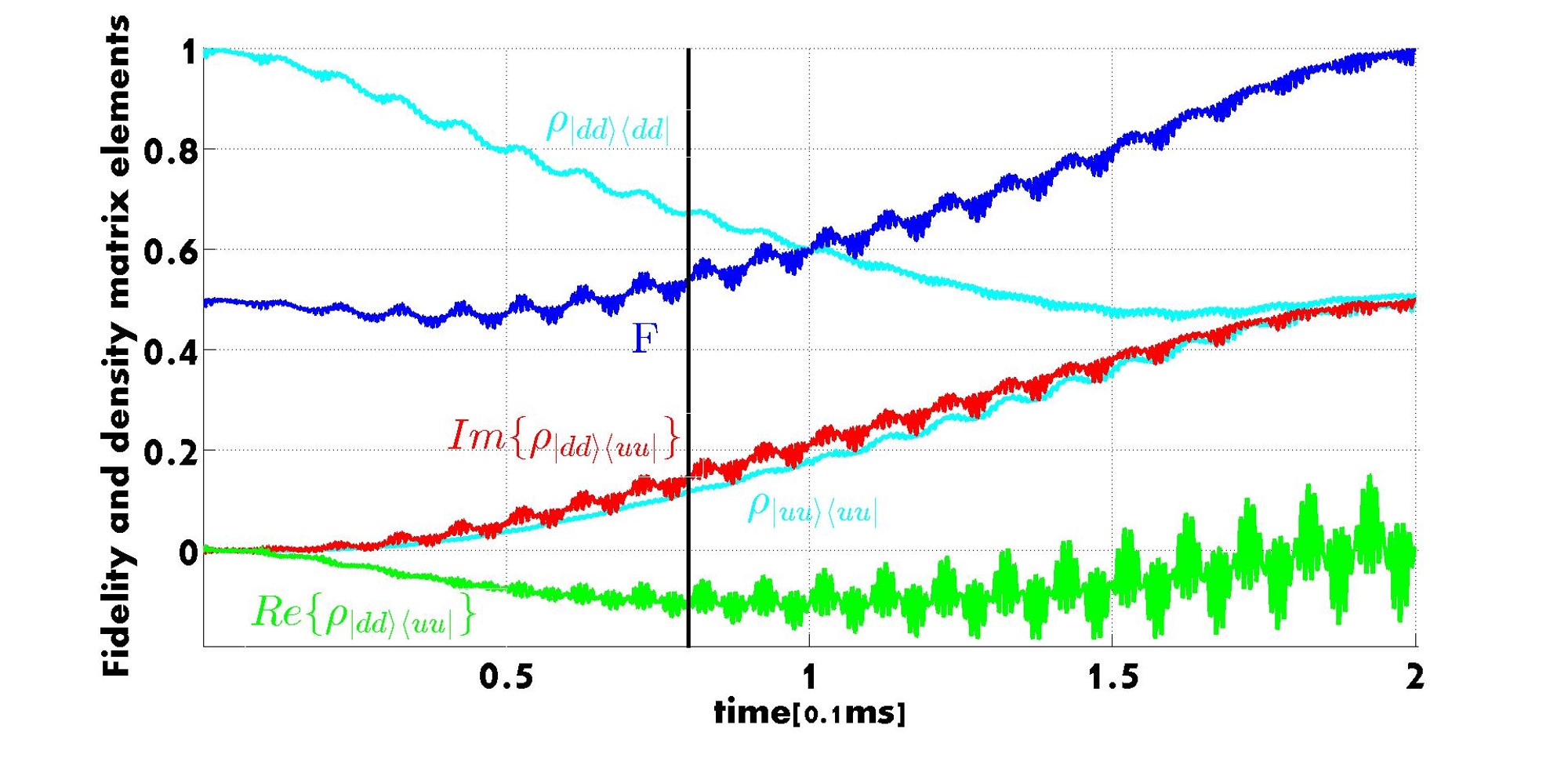}
  \caption{{\bf Entangling gate performance.} The dynamics of two qubit gate starting with $\left\vert dd\right\rangle$ as the initial state with the ground state of the vibration mode is plotted.  We obtain the maximum entangled state $\left\vert \Psi_+\right\rangle=(\left\vert dd\right\rangle+i\left\vert uu\right\rangle)/\sqrt{2}$ at the end of the gate,   after applying a $\pi$-phase flip of the RF driving field in the middle of the gate. Counting from above at $time=0.08msec$, the first (cyan) curve is $\rho_{\left\vert dd\right \rangle \left\langle dd \right\vert}$, the second (blue) is the fidelity of being in the maximum entangled state $F= \left\langle\Psi_+\right\vert\rho\left\vert \Psi_+\right\rangle$, the third (red) is the imaginary part of $\rho_{\left\vert dd\right \rangle \left\langle uu \right\vert}$, the fourth (cyan) is $\rho_{\left\vert uu\right\rangle \left\langle uu \right\vert}$, and the lowest one (green) is the real part of $\rho_{\left\vert dd\right \rangle \left\langle uu \right\vert}$.}
   \label{performance}
\end{figure}

We can evaluate the higher order contribution of the neglected terms using the simulation of the gate performance, including all the terms mentioned above
without any approximation. In the simulation we have two Yb ions of mass $m=173amu$, magnetic gradient $\partial_z B =65T/m$, secular frequency $\nu=2\pi\cdot500kHz$, effective Lamb-Dicke parameter $\eta=0.01$, microwave field Rabi frequency $\Omega=2\pi\cdot 495kHz$, RF field Rabi frequency $\Omega_r=2\pi\cdot99kHz$, one loop in phase space $k=1$, the off-resonant microwave driving field contribution where $\Delta\omega_0=2\pi\cdot 5MHz$, starting in the vibrational ground state, namely, the initial number of phonons is $N=0$. This gives a gate duration of $t_{gate}=0.2$ msec. In addition, we applied the $\pi$-phase flip in the middle of the gate (fig. \ref{performance}).

\begin{figure}
   \centering
  \includegraphics[width=0.5\textwidth]{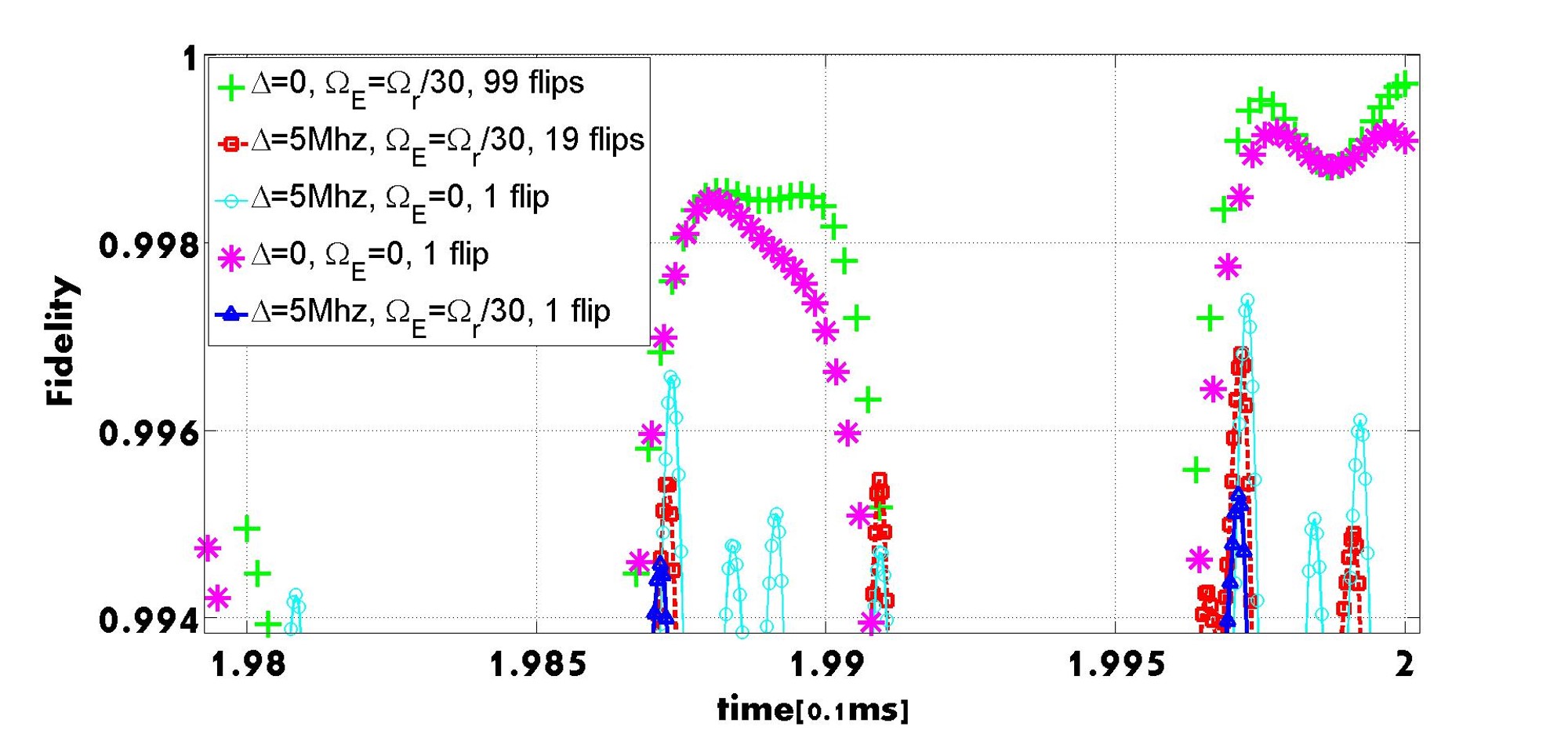}
  \caption{{\bf Zooming-in of the gate fidelity (eq. \ref{performance}) assuming different terms in the Hamiltonian.} From the comparison between the achieved fidelity of the case where the electric field and the single addressing vanish (pink asterisks), to the case where only the single addressing is taken into consideration (cyan circles), we understand that the infidelity due to the single addressing is $IF=0.002$. Without any assumption, while applying only a single $\pi-$phase flip in the middle of the gate (blue triangles), the gate fidelity is $F=0.995$. By increasing the number of  $\pi-$phase flips the fidelity is increased as-well. In the case where we apply 19 $\pi-$phase flips during the gate (red rectangular) the achieved fidelity almost reaches $ F=0.997$. When we suppress the single addressing (in future configurations) and apply 99 $\pi-$phase flips (green pluses) the infidelity is $IF=3e^{-4}$, almost reaching the fault tolerance threshold.  }
   \label{fidelity}
\end{figure}

We find that the higher order expansion of eq. \ref{delta}, is an even function of $\Omega_r$, which is not refocused by the spin-echo. It yields infidelity of $IF=0.002$ (fig. \ref{fidelity}) 
. However, although the current set-up does not allow it, in future implementations, the ions may be aligned along a perpendicular direction to the magnetic gradient, such that the ions would feel the same magnetic field, and thus have the same bare energy structure due to Zeeman splitting. Therefore, the undesired contribution of eq. \ref{delta} would vanish. Simulating all the terms but eq. \ref{delta} yields infidelity of $IF=3\cdot 10^{-4}$ (fig. \ref{fidelity}), getting closer to the fault tolerance threshold.

For these suggested parameters, the contributions of the coupling terms, eq. \ref{XY1} and \ref{ZZ1}, correspond to infidelity of $1.2\cdot 10^{-5}$ and $5\cdot 10^{-5}$ respectively. 


\section{robustness to noise}
There are several noise sources in this proposal, which are significantly suppressed using concatenated continuous dynamical decoupling  \cite{DD6}. The main noise source originates from ambient magnetic field fluctuations. This is suppressed by the resonant microwave driving field \cite{gate realization WUNDERLICH1,WINNI}, which can be readily shown in the dressed state basis, the eigenstates of $\sigma_x$. The $x$ direction is orthogonal to the quantization axis, and therefore to the magnetic noise which is described by the operator $\sigma_z$ (fig. \ref{bare}); thus, by moving to the interaction picture of the dressed state energy structure, as was done in eq.\ref{red}, the magnetic noise is suppressed in the RWA, exactly as is the case in the gate scheme proposed by Bermudez {\it et al.} \cite{gate proposal BERMUDEZ1}.


The second important noise source originates from fluctuations in the Rabi frequency of the resonant microwave driving field. The Rabi frequency fluctuations are one of the most fidelity damaging sources in reference \cite{gate realization BERMUDEZ}. To protect our gate from this, we use a resonant RF or a detuned microwave driving field (eq.\ref{radio2}), whose effect can be shown in the double-dressed state basis, the eigenstates of $\sigma_z$. The $z$ direction is orthogonal to the resonant microwave driving field's polarization, and therefore to the Rabi frequency fluctuations, which is described by the operator $\sigma_x$ (fig. \ref{dressed RF}, \ref{dressed M}); thus, by moving to the interaction picture of the double-dressed state energy structure, as was done in eq.\ref{red2}, the suppression of the Rabi frequency noise is achieved. Since the Rabi frequency of the resonant microwave field is much larger than the Rabi frequency of the RF or the detuned microwave driving field, the noise originated from fluctuations in one of those driving fields is less fidelity-damaging.

We can estimate the infidelity due to these noise sources. Previous experiments \cite{gate realization WUNDERLICH1,WINNI} have shown that the dephasing rate of the ambient magnetic field noise was suppressed by the dressing microwave radiation to less than $S_{B,B}(20kHz)=1$ Hz. With higher dressing Rabi frequencies the dephasing rate would be decreased as 
\begin{equation}
S_{B,B}(\Omega)=S_{B,B}(20kHz)\left(\frac{20kHz}{\Omega} \right)^2=0.0016Hz,
\end{equation}
assuming that the correlation time of the magnetic noise is much longer than $50\mu sec$.

The dephasing rate due to the Rabi frequency fluctuations without additional dressing fields can be approximated as $S_{\Omega,\Omega}(0)\approx 0.01 \Omega$. The time correlation of this noise source is long in most experiments, i.e., of the same order as the two qubit gate rate. Thus, by applying the second dressing fields, the first order noise is completely suppressed $S_{\Omega,\Omega}(\Omega_r)\approx 0$. Yet, the second order contribution of this noise, which is an A.C Stark shift, becomes dominant and behaves as $  S_{\Omega,\Omega}(0)^2/\Omega_r$. The second order contribution of the Rabi frequency noise source has a long time correlation, and since it is an odd function of $\Omega_r$, with the spin-echo $\pi-$phase flip, its contribution is negligible.

The dephasing rate due to the Rabi frequency of the last dressing field, the RF or the detuned microwave driving field, can be approximated as before as $S_{\Omega_r,\Omega_r}(0)\approx 0.01\Omega_r$. We can assume that the time correlation of this noise source is very long, therefore, it can be refocused using the spin-echo.

In total, the infidelity caused by the ambient magnetic field and the fluctuations of the first and second dressing fields can be estimated as $IF=3\cdot 10^{-7}$ for the $0.2msec$ gate duration. This is much smaller than the fault tolerance threshold, and therefore the gate is robust to the main noise sources.


High fidelity depends on good timing (fig. \ref{fidelity}). 
In this regard, techniques such as pulse shaping as were suggested by C. F. Roos \cite{gate proposal ROOS}, and Hayes et al. \cite{pi} have been developed in order to overcome this problem. Yet, these techniques can not be implemented in the proposed scheme for the microwave-based implementations, since we can not modify the amplitude of the gate operator $(\nu\eta/4)$ (eq. \ref{red2}), as these techniques require. Nevertheless, microwave signals can be created using highly accurate arbitrary waveform generators which can address this timing issue. 

In addition, timing depends also on the stability of the Rabi frequency of the resonant microwave $\Omega$ and secular frequency $\nu$ (eq. \ref{Agate}). The fluctuations in the Rabi frequency are suppressed by the second dressing fields, and as is estimated at \cite{lucas}, the motional error of should be less then $10^{-3}$ in a $0.2msec$ gate duration. 

\section{undesired effect of the RF driving field}
Applying a driving field closely-detuned from the selected mode of motion coherently increases the number of phonons in the system. Although the driving field is magnetically coupled to the spin, its associated electric field is coupled directly to the motion, and causes displacement of the ion motion. This interaction can be written as 
\begin{equation}
H_{d}=eE \cdot z = \Omega_E\left( 
b^\dagger + b \right)\cos (\nu-\epsilon)t,
\end{equation}
where $\Omega_E=\frac{eE_z}{\sqrt{2 m \nu}}$ and $e$ is the ion charge.
Fortunately, the detuning $\epsilon $ reduces these effects, and protects against excessive motional displacement. In the motion's interaction picture, the displacement becomes $H_{{d}_I}= \frac{\Omega_E}{2}\left( b^\dagger e^{i \epsilon t} + h.c \right) $, after the RWA, when assuming $\Omega_E \ll 4\nu$.  Since the gate Hamiltonian and the additional undesired displacement term have the same rotating exponents, the Hamiltonian of both of them can be represented by 
\begin{equation}
H_{gate}=\left( \frac{\eta\nu }{4}\sum_j \sigma^j_z+\frac{ \Omega_E}{2} \right) \left( b^\dagger e^{i\epsilon t} +b e^{-i\epsilon t} \right),
\end{equation}
where, as before, an exact time propagator exists. Completing an integer number of circles in phase space results in the desired phase gate transition with a correction which can be treated by moving to the interaction picture, or by either a $\pi$-pulse or a $\pi$-phase flip after half of the gate time, as was discussed above. 

The real influence of the electric coupling is the number of phonons that are coherently added during the gate operation, before closing a circle in phase space. It can be easily understood in the representation of the Magnus expansion. The first order of the Magnus expansion (eq. \ref{magnus}) is the displacement operator, which is maximized in half of a cycle in phase space. For example,  at $t=\tau/ 2K$, the displacement operator is $D(\alpha(\frac{\tau}{2K}))=e^{\alpha(\frac{\tau}{2K}) b^\dagger - h.c}$, where $\alpha(\frac{\tau}{2K})=  \frac{\sum_j\sigma^j_z }{2 \sqrt{K}}+\frac{ \Omega_E}{\epsilon} $. If we initialize our system in the motional ground state, after $t=\tau/ 2K$, the motional state becomes the coherent state $\left \vert \psi\right\rangle_{\frac{\tau}{2K}} = \left\vert \alpha(\frac{\tau}{2K}) \right\rangle $, whose average number of phonons could reach $<N>=\left\vert \alpha(\frac{\tau}{2K}) \right\vert^2 \sim (\Omega_E/\epsilon)^2 $, if $ \epsilon \ll\Omega_E$. If there are too many phonons in the system important assumptions such as the Lamb-Dicke assumption ($\nu\eta\sqrt{N}/4 \ll \Omega_r \pm \epsilon $) made in eq. \ref{Lamb-Dicke regime} would not hold and lead to a reduction in the achievable fidelity.

It is therefore advantageous to ensure that $\Omega_E \ll \epsilon$, such that the electric coupling can be adiabatically eliminated in the second-order perturbation approach. To fulfil the restrictions in eq. \ref{assumption} ($\epsilon \ll \Omega_r$), the ratio between the magnetic coupling and the electric coupling $\frac{\Omega_r}{\Omega_E}$ must be large. If it is not the case, and the electric field is large, a sequence of many $\pi-$phase flips can be applied in order to suppress the population of high phonon states. Note that this sequence can be performed even within a single cycle in phase plane $(K=1)$.

In the near field regime, the magnetic field dominates over the electric field by a factor of the wavelength divided by the size of the system. Therefore, the ratio between the magnetic coupling and the electric coupling can be estimated in the following way. For a RF field of frequency $at least one of the spin states should beKHz$ and dimension $1cm$, the typical velocity of the electrons is $3\times 10^{-6} c,$ where $c$ is the speed of light. In that case the electric term in the Hamiltonian would be $\Omega_E=e v \times B \times x_0,$ where $e$ is the ion charge, $B$ the magnetic field and $x_0$ is the standard deviation of the wave function which is of the order on $10nm.$ Thus, the ratio between the magnetic and the electric energy is $\frac{\Omega_r}{\Omega_E}=\frac{g\mu_B B}{e v B x_0} \approx 30.$ By reducing the dimension of the system a better reduction factor can be achieved. It means that even in the worst case, when lacking the ability to set the polarization of the RF driving field along the axial axis, the gate could still be performed with high fidelity. 

In addition to the above, unwanted E-fields can be created due to the finite resistance of a coil which creates the required RF field. In order to investigate this we simulated a coil that consists of 3 turns of copper wire with a wire diameter of 6.7 mm and a coil diameter of 4 cm. The coil is placed 1 cm from the ions. To make this even more accurate we have included the two feed-lines going to the coil. The E-field along the trap axis created due to the potential drop across the coils operated at a frequency of 500 kHz and a current sufficient to obtain $\Omega_r = 2\pi \cdot 100$ kHz has been found to be $2.04 \cdot 10^{-5}$ V/m, which is negligible compared to the Lorentz force discussed above.   

Adding the rotating electric field into the simulation of the gate performance, with $\Omega_E=\Omega_r/30$, yields infidelity of $IF=5\cdot 10^{-3}$ after a single spin-echo (fig. \ref{fidelity}). However, assuming that the two ions have the same bare energy gap (such that eq. \ref{delta} can be dropped), after a sequence of $19$ and $99$ phase flips, the infidelity reduces to $3\cdot 10^{-3}$ and $3\cdot 10^{-4}$ respectively  (fig. \ref{fidelity}).



\section{Moving to the interaction pictures}
In our proposal we move three times to different interaction pictures, with respect to: (1) the bare energy structure $\sum_j(\omega^j_0/2) \sigma^j_z$, (2) the dressed state energy structure  $(\Omega/2) \sum_j\sigma^j_x$, and (3) the double dressed state energy structure  $(\Omega_{r}/2)\sum_j \sigma^j_z$. The way it is done in the lab is the following. The driving fields accumulate the phases corresponding to the first interaction picture with respect to the bare energy structure, namely, the system automatically moves to the first interaction picture. 

As regards to moving to the second interaction picture with respect to the dressed state energy structure, at the end of the experiment (after time $\tau$), after eliminating the RF field, we operate only with the microwave dressing fields for additional time $t_{add}$, during which the accumulated phase of the second interaction picture trivially vanishes, namely $\Omega(\tau+t_{add})/2=2\pi n$ for an integer $n$. However, since we cannot shut down the static magnetic field gradient, we have to change the microwave Rabi frequency ($\Omega \rightarrow \Omega_{new}$), such that it would be far detuned from the secular frequencies, thus adiabatically eliminating the magnetic gradient's contribution. Now the condition should be $(\Omega/2)\tau+(\Omega_{new}/2)t_{add}=2\pi n$ for an integer $n$.

The third interaction picture with respect to the double dressed energy structure, is carried via the $\pi-$phase flip of the RF driving field.

\section{realization with lasers}
Until now, we have discussed the realization of the gate in a microwave-based implementation. However, with a similar derivation, our scheme could also be implemented in laser-based set-ups. Dealing with lasers, the spin to phonon coupling is carried via the laser's sufficient large momentum, $k$. Therefore, the ion's two spin states no longer have to be sensitive to the magnetic field. Here, we may use trapped ions, 
with a qubit transition frequency in the optical, or even in the microwave regime, using resonant or Raman transitions, respectively. 

\begin{figure}
   \centering
  \includegraphics[width=0.5\textwidth,natwidth=1280,natheight=720]{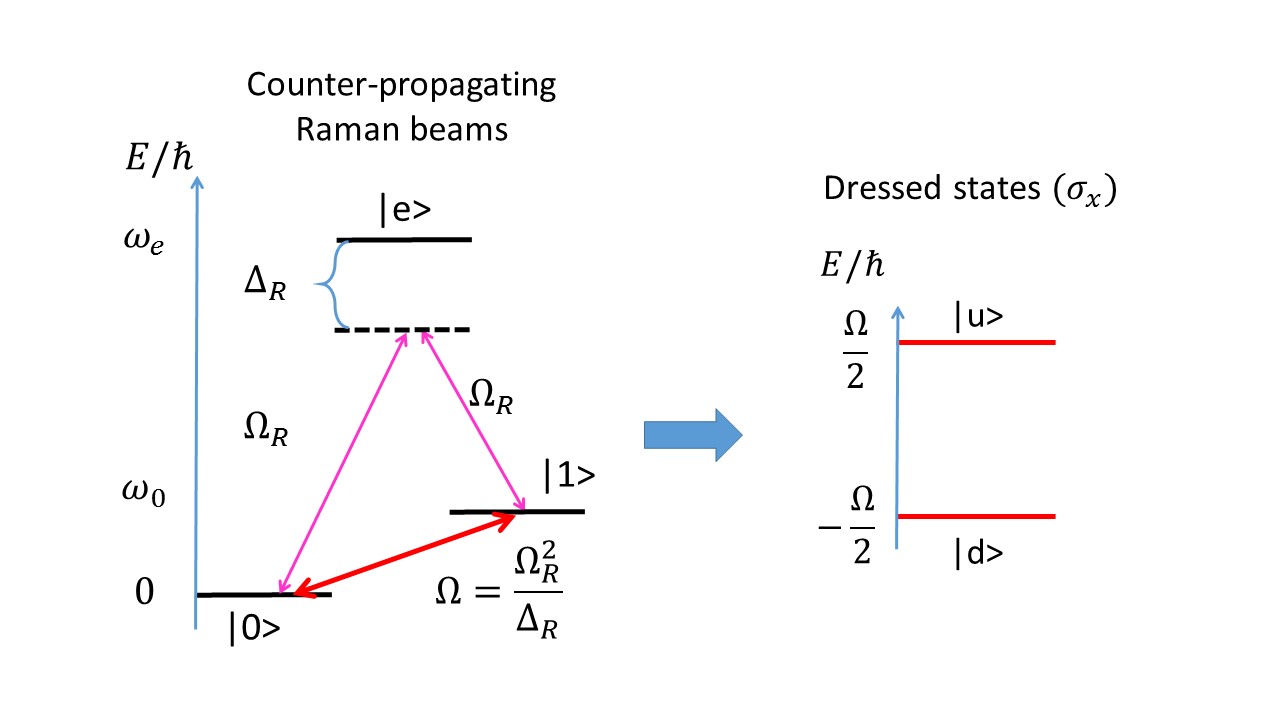}
  \caption{{\bf Moving from the bare states to the dressed states using Raman beams.} 
  In the laser-based designs, if the qubit transition frequency is in the microwave regime, by applying counter-propagating Raman beams that effectively generate a resonant transition of the bare state basis, we move to the dressed state basis, in a perpendicular direction $\frac{\Omega}{2}\sum_j\sigma_x^j$. We demand the counter-propagation of the beams in order to obtain the coupling to the phonons, which is needed for the entangling gate.
   }
    \label{counter Raman}
\end{figure}

After the RWA, where we assume that the Rabi frequency of the resonant laser transition (fig. \ref{bare}), or the effective Rabi frequency of the counter-propagating Raman beams (fig. \ref{counter Raman}), is much lower than the qubit energy splitting, $\Omega \ll 4\omega_0$, the Hamiltonian of the ion motion and the laser interaction in the rotating frame of the bare state energy structure $(\omega_0/2)\sigma_z$ is given by 
\begin{equation}
\begin{split}
H_I=\nu b^\dagger b+\frac{\Omega}{2}\sum_j \left( \sigma^j_+ e^{i\eta_L( b^\dagger +b) }+h.c \right), \\
\end{split}
\label{start laser}
\end{equation}
where $\eta_L= k Z /\sqrt{2 m \nu}$ is the Lamb-Dicke parameter, with $Z$ being the normal mode coefficient. 

Performing the experiment in the Lamb-Dicke regime, where $\eta_L\sqrt{N} \ll 1$, and $N$ is the number of phonons in the system, enables one to expand the Lamb-Dicke exponent to the second order in $\eta_L$. Thus, we obtain a similar Hamiltonian to the Hamiltonian of the microwave-based implementation (eq.\ref{start}) given by
\begin{eqnarray}
H_I&=&\nu b^\dagger b \nonumber \\ 
&+&\frac{\Omega}{2} \sum_j\left\lbrace \sigma^j_x + \eta_L\sigma^j_y\left(   b^\dagger +b  \right) - \frac{\eta_L^2}{2}\sigma^j_x\left(   b^\dagger +b  \right)\left(   b^\dagger +b  \right)\right\rbrace, \nonumber \\
\label{start laser2}
\end{eqnarray}
where the carrier transition performs the dynamical decoupling, the sideband terms would perform the gate, and the additional term is an undesired phonon dephasing term.

\begin{figure}
   \centering
  \includegraphics[width=0.5\textwidth,natwidth=1280,natheight=720]{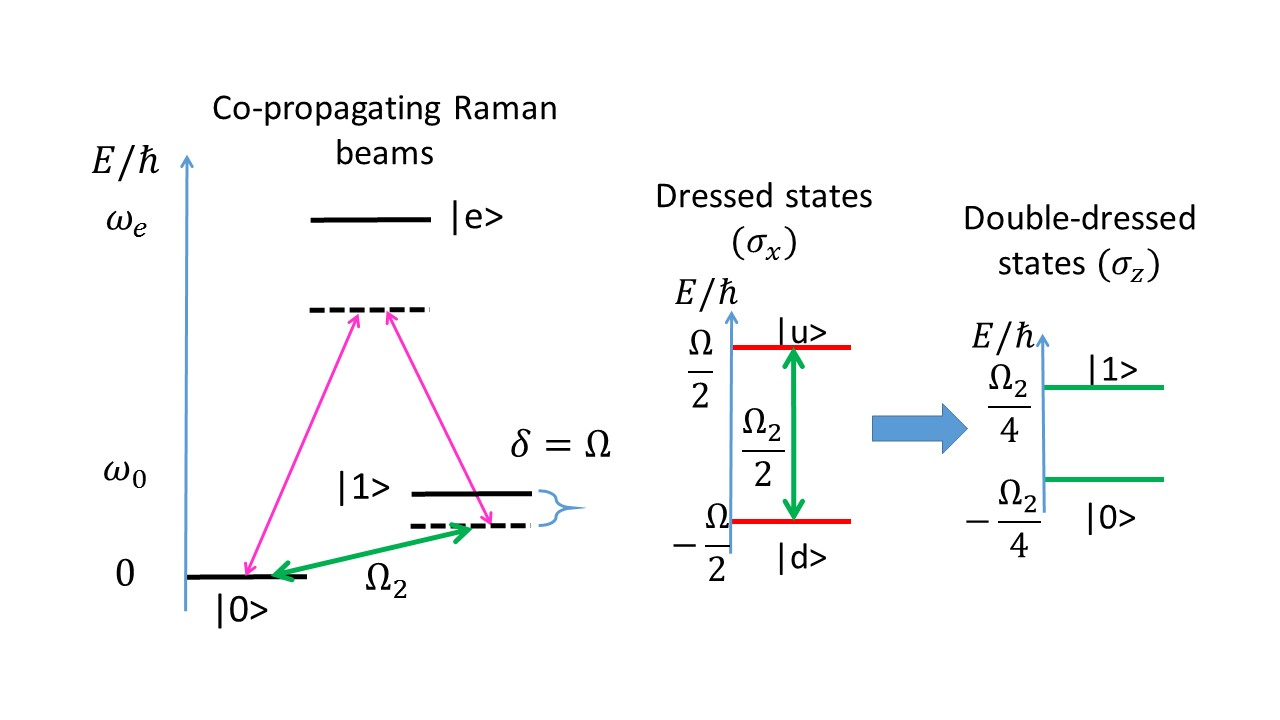}
  \caption{{\bf Moving from the dressed states to the double dressed states using co-propagating Raman beams to generate a detuned transition in the bare state basis.} In the laser-based designs, if the qubit transition frequency is in the microwave regime, applying another co-propagating Raman beams that generate the detuned transition in the base state basis, such that $\delta=\Omega$, is effectively equivalent to driving resonant transition with the dressed state energy structure. This takes us to the double-dressed state basis, in a perpendicular direction $\frac{\Omega_2}{4}\sum_j\sigma_z^j$. The energy gap in the perpendicular direction protects against the noise that originates from the Rabi frequency fluctuations $\Omega$.  
  }
    \label{co Raman}
\end{figure}

After transforming to the dressed state basis, and moving to the interaction picture with respect to the dressed state energy splitting, 
the sideband transitions, similar to those from eq.\ref{red} in the equivalent microwave case, are observed with the undesired phonon-dephasing term \cite{plenio1,plenio2}. This is given by
\begin{equation}
\begin{split}
H_{red}=-i \frac{\eta\Omega}{2} \left(  \sum_j S^j_+ e^{i\Omega t} -h.c \right) \left(  b^\dagger e^{i\nu t} +h.c   \right)\\ 
 - \frac{\eta^2\Omega}{4}\sum_jS^j_z \left(  b^\dagger e^{i\nu t} +h.c   \right)\left(  b^\dagger e^{i\nu t} +h.c   \right).
\label{red laser}
\end{split}
\end{equation}
Transforming eq. \ref{red laser} to the desired gate transitions, while suppressing the dephasing term, can be achieved in several ways. One could apply a resonant RF driving field (fig. \ref{dressed RF}), as was done in the microwave based design discussed above. If the bare state energy splitting is in the microwave regime one can apply an additional detuned microwave driving field (fig. \ref{dressed M}), which can alternatively be realized by applying co-propagating Raman lasers to perform the detuned transition as shown in fig. \ref{co Raman}. If the bare state energy splitting is in the optical regime, we could use a detuned laser driving field, whose additional spin-phonon coupling terms are adiabatically suppressed (fig. \ref{dressed M}). For simplicity, we consider the case of applying an RF driving field given by $H_{rf}=\Omega_r\sum_j\sigma^j_z\cos \Omega t$, as was done in the microwave based scheme described above.

In the interaction picture according to the dressed state energy structure, the RF driving field becomes $(-\Omega_{r}/2) \sum_jS^j_x$, as shown in eq.\ref{radio2}(a). Transforming to the double-dressed state basis, and moving to the interaction picture with respect to the new energy structure, the noise, which comes from fluctuations in the Rabi frequency induced by the laser driving fields, is suppressed and we are left with the phase gate in the double-dressed state basis, or the MS gate in the dressed state basis which is given by
\begin{equation}
H_{gate}=i\frac{\eta_L\Omega}{4}\sum_j\sigma^j_z(b^\dagger e^{i\epsilon t} + h.c)=-i\frac{\eta_L\Omega}{4}\sum_jS^j_x(b^\dagger e^{i\epsilon t} + h.c)
\label{red2 laser}
\end{equation}
respectively and is equivalent to the microwave-based case.

\section{summary}
We have presented a multi-qubit gate proposal for trapped ions which can be used in both, microwave-based and laser-based, implementations. The gate is robust to the main fidelity damaging noise sources (ambient magnetic field fluctuations and Rabi frequency fluctuations) and is decoupled from phonons. Our proposal builds on the scheme proposed by Lemmer {\it et al.} \cite{gate proposal BERMUDEZ2}, however, our gate can be implemented using microwave as well as laser based experimental setups. The main difference is that in our proposal the gate transitions originate from the resonant driving field, such that we can use the carrier transition for performing continuous dynamical decoupling, instead of generating it with an additional driving field. We use a second driving field for both, transforming the red-sideband transitions to the desired gate transitions and for continuously applying additional dynamical decoupling to suppress noise from Rabi frequency fluctuations, which originate from the first resonant driving field. 

We thank Nati Aharon for useful discussions. We acknowledge the support of the European commission (STReP EQUAM).  
Moreover, this work has benefitted from financial support by the COST Action MP1001 "Ion Traps for Tomorrow's Applications". This work is supported by the U.K. Engineering and Physical Sciences Research Council  [EP/G007276/1, the UK Quantum Technology hub for Networked Quantum Information Technologies (EP/M013243/1), the UK Quantum Technology hub for Sensors and Metrology (EP/M013243/1)], the European Commission’s Seventh Framework Programme (FP7/2007-2013) under Grant Agreement No. 270843 (iQIT), the Army Research Laboratory under Cooperative Agreement No. W911NF-12-2-0072, the US Army Research Office Contract No. W911NF-14-2-0106 and the University of Sussex. The views and conclusions contained in this document are those of the authors and should not be interpreted as representing the official policies, either expressed or implied, of the Army Research Laboratory or the U.S. Government. The U.S. Government is authorized to reproduce and distribute reprints for Government purposes notwithstanding any copyright notation herein.


\end{document}